\documentclass[acmsmall]{acmart}

\newif\ifextended
\extendedtrue

\AtBeginDocument{%
  \providecommand\BibTeX{{%
    \normalfont B\kern-0.5em{\scshape i\kern-0.25em b}\kern-0.8em\TeX}}}

\setcopyright{none}
\copyrightyear{2018}
\acmYear{2018}
\acmDOI{10.1145/1122445.1122456}

\acmJournal{PACMPL}
\acmVolume{POPL}
\acmNumber{1}
\acmArticle{1}
\acmMonth{1}

\usepackage{algorithm}
\usepackage[noend]{algpseudocode}
\usepackage{amsmath}
\usepackage{stmaryrd}
\usepackage{url}
\usepackage{listings}
\usepackage{mdwlist}
\usepackage{pifont}
\usepackage{graphicx}
\usepackage{wrapfig}
\usepackage{fancyvrb}
\usepackage{paralist}
\usepackage{caption}
\usepackage{comment}
\usepackage{bbm, dsfont}
\usepackage{mathrsfs}
\usepackage{xcolor}
\usepackage{tikz} \usetikzlibrary{patterns}
\usepackage{pgfplots}
\usepackage{esvect}
\usepackage{xspace}
\usepackage{array, multirow}
\usepackage{rotating, makecell}
\usepackage{enumitem}

\definecolor{codegreen}{rgb}{0,0.6,0}
\definecolor{codegray}{rgb}{0.5,0.5,0.5}
\definecolor{codepurple}{rgb}{0.58,0,0.82}
\definecolor{backcolour}{rgb}{1,1,1}

\lstdefinestyle{mystyle}{
    xleftmargin=2em,
    backgroundcolor=\color{backcolour},
    commentstyle=\color{codegreen},
    keywordstyle=\color{red},
    numberstyle=\tiny\color{codegray},
    stringstyle=\color{codepurple},
    basicstyle=\ttfamily\small,
    breakatwhitespace=false,
    breaklines=true,
    captionpos=b,
    keepspaces=true,
    numbers=left,
    numbersep=5pt,
    showspaces=false,
    showstringspaces=false,
    showtabs=false,
    tabsize=2
}

\newcommand{\irule}[2]%
   {\mkern-2mu\displaystyle\frac{#1}{\vphantom{,}#2}\mkern-2mu}
\newcommand{\irulelabel}[3]
{
\mkern-2mu
\begin{array}{ll}
\displaystyle\frac{#1}{\vphantom{,}#2} & #3
\end{array}
\mkern-2mu
}

\newcommand{\toolname}{\textsc{NetRep}\xspace}
\newcommand{\jaid}{\textsc{Jaid}\xspace}

\newcommand{\prog}{\mathcal{P}}
\newcommand{\sketch}{\Omega}
\newcommand{\grammar}{\mathcal{G}}
\newcommand{\class}{\mathcal{C}}
\newcommand{\func}{F}
\newcommand{\lineid}{L}
\newcommand{\stmt}{s}
\newcommand{\imm}{v}
\newcommand{\term}{\psi}
\newcommand{\addr}{\delta}
\newcommand{\cstr}{\Phi}

\newcommand{\lineMap}{\mathcal{B}}
\newcommand{\pathMap}{\pi}
\newcommand{\memMap}{M}
\newcommand{\exs}{\mathcal{E}}
\newcommand{\summary}{\mathcal{S}}
\newcommand{\visited}{\mathcal{V}}
\newcommand{\abs}{\mathcal{A}}
\newcommand{\hole}{??}

\newcommand{\denot}[1]{\llbracket #1 \rrbracket}
\newcommand{\set}[1]{\{ #1 \}}
\newcommand{\dom}{\emph{dom}}

\newcommand{\newStmt}[2]{#1 := \textbf{new}~ #2}
\newcommand{\assignStmt}[2]{#1 := #2}
\newcommand{\jmpStmt}[2]{\textbf{jmp} ~(#1)~ #2}
\newcommand{\staticCallStmt}[4]{#1 := #2.#3(#4)} 
\newcommand{\virtualCallStmt}[4]{#1 := #2.#3(#4)} 
\newcommand{\retStmt}[1]{\textbf{ret}~ #1}

\begin{document}

\title{NetRep: Automatic Repair for Network Programs}
\author{Lei Shi}
\affiliation{
  \institution{University of Pennsylvania}
  \city{Philadelphia}
  \state{PA}
  \country{USA}
}
\author{Yuepeng Wang}
\affiliation{
  \institution{Simon Fraser University}
  \country{Canada}
}

\author{Rajeev Alur}
\affiliation{
  \institution{University of Pennsylvania}
  \city{Philadelphia}
  \state{PA}
  \country{USA}
}
\author{Boon Thau Loo}
\affiliation{
  \institution{University of Pennsylvania}
  \city{Philadelphia}
  \state{PA}
  \country{USA}
}

\begin{abstract}
Debugging imperative network programs is a challenging task for developers because understanding various network modules and complicated data structures is typically time-consuming. To address the challenge, this paper presents an automated technique for repairing network programs from unit tests. Specifically, given as input a faulty network program and a set of unit tests, our approach localizes the fault through symbolic reasoning, and synthesizes a patch such that the repaired program can pass all unit tests. It applies domain-specific abstraction to simplify network data structures and utilizes modular analysis to facilitate function summary reuse for symbolic analysis. We implement the proposed techniques in a tool called \toolname and evaluate it on 10 benchmarks adapted from real-world software-defined networking controllers. The evaluation results demonstrate the effectiveness and efficiency of \toolname for repairing network program.
\end{abstract}

\maketitle

\section{Introduction} \label{sec:intro}

Emerging tools for program synthesis and repair facilitate automation of programming tasks in various domains. For example, in the domain of end-user programming, synthesis techniques allow users without any programming experience to generate scripts from examples for extracting, wrangling, and manipulating data in spreadsheets~\cite{flashfill-popl11,flashmeta-oopsla15}. In computer-aided education, repair techniques are capable of providing feedback on programming assignments to novice programmers and help them improve programming skills~\cite{Wang-pldi18,Gulwani-pldi18}. In software development, synthesis and repair techniques aim to reduce the manual efforts in various tasks, including code completion~\cite{completion-pldi14,codehint-icse14}, application refactoring~\cite{Raychev-oopsla13}, program parallelization~\cite{par-pldi17}, bug detection~\cite{LeGoues-cacm19,deepbugs-oopsla18}, and patch generation~\cite{LeGoues-cacm19,prophet-popl16}.

As an emerging domain, Software-Defined Networking (SDN) offers the infrastructure for monitoring network status and managing network resources based on programmable software, replacing traditional specialized hardware in communication devices. 
Since SDN provides an opportunity  
 to dynamically modify the traffic handling policies on programmable routers, this technology has witnessed growing industrial adoption.
 However, using SDNs
involves many programming tasks that are inevitably susceptible to programmer errors, thereby introducing routing bugs~\cite{vericon-pldi14,veriflow-nsdi13}. For example, a device with incorrect routing policies could forward a packet to undesired destinations, and  a buggy firewall rule may make the entire network system vulnerable to security threats.

Debugging network programs is a challenging task. First, it is difficult and time-consuming to pinpoint the fault location. Even if some test discovers the behavior of a network module is different from expected, it does not provide more detailed information about the fault location. Developers still need to go through a large number of functions to locate the root cause of the problem. Second, fixing the fault requires developers to have a good understanding of many modules with which developers are not necessarily familiar, even though the final patch only contains a few lines of code. Third, when bugs manifest in production systems, given high volume of network traffic and potential packet reordering, bugs may not be deterministic and reproducible.

To help developers with programming of SDN, several synthesis and repair techniques are proposed in recent years. For instance, McClurg et al.~\cite{McClurg-pldi15,McClurg-cav17} have developed synthesis techniques for automatically updating global configurations and generating synchronizations for distributed controllers from high-level specifications. In addition, prior work~\cite{MetaProvenance-hotnets15,MetaProvenance-nsdi17,McClurg-fmcad16} presents automated repair technique for fixing network policies written in declarative domain-specific languages such as Datalog and horn clauses. 

While existing techniques have shown the promise of diagnosing and ensuring correctness of network programs written in declarative DSLs, they are not applicable to many popular network controllers (e.g. Floodlight~\cite{floodlight-web21}) written in imperative languages such as Java and C++ due to the complex language features.
Furthermore, general-purpose repair tools are not mature enough to be applied  directly to network programs. 
Our insight is that network controllers are typically divided into layers and organized as several modules, which presents good opportunities for repair techniques to achieve better scalability. However, general repair tools are unaware of the code structure of network programs and thus cannot leverage it for optimization.

Motivated by the demand of automated repair and the limitations of existing techniques, we develop a precise and scalable program repair technique for network programs. Specifically, our repair technique takes as input a network program and a set of unit tests, reveals the program location that causes the test failure, and automatically generates a patch to fix the program.
Developers only need to provide unit tests that reveal the fault but do not need to manually investigate the implementation or write a patch. Thus, our technique can help developers significantly reduce the burden of debugging network programs.

\begin{figure}[t]
\includegraphics[scale=0.7]{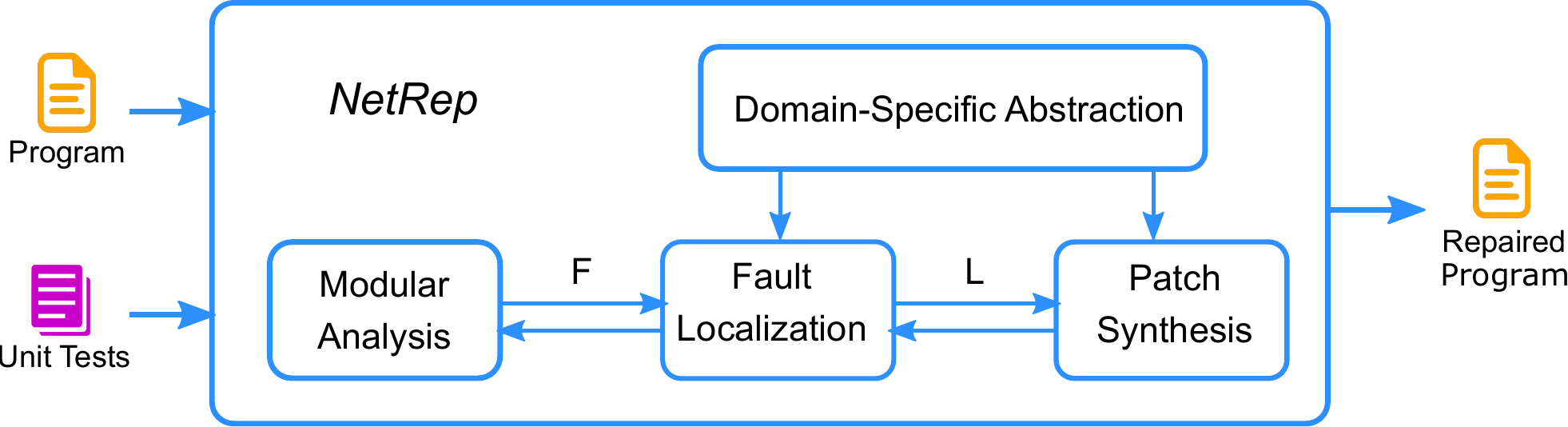}
\caption{Schematic workflow.}
\label{fig:workflow}
\vspace{-10pt}
\end{figure}

Figure~\ref{fig:workflow} presents a schematic workflow of our approach to repairing network programs. In summary, given a network program to repair and a set of unit tests, our approach consists of the following techniques:
\begin{itemize}[leftmargin=*]
\item \emph{Domain-specific abstraction.}
Unlike general program repair techniques, our repair approach is specialized in network programs and leverages a high-level specification of network modules as the domain-specific abstraction to enable scalable analysis. Utilizing the abstraction allows us to reason about the behavior of various network data structures without going through complex implementations.
\item \emph{Modular analysis.}
In order to scale to large network programs, we take a modular fault localization and repair approach. Specifically, we divide the network program into smaller functions and focus on repairing one function at a time. This enables reuse of function summaries, which significantly simplifies the symbolic reasoning and thus makes our repair techniques more scalable.
\item \emph{Symbolic fault localization.}
Given a network program and failing test cases, we obtain a set of SMT formulas through symbolic execution, which encode the semantics of programs and tests. Since some test cases are failed, the obtained SMT formulas are not consistent with each other, i.e. the conjunction is unsatisfiable. To pinpoint the fault location, we identify a large subset of formulas by relaxing the original formulas with boolean guards. Intuitively, the model of boolean guards for relaxation indicates the potential fault localization, because it suggests changing corresponding parts of the program can make the program consistent with the provided tests.

\item \emph{Patch synthesis.}
After identifying the potential fault location, we perform program synthesis to automatically generate a patch from the test cases. In particular, our synthesis technique constructs a context-free grammar to capture the search space based on the context (e.g., variables and functions in scope) of the fault location and conducts enumerative search to find a patch that is generated by the grammar so that the corresponding patched program can pass all unit tests.
\end{itemize}

Based on these ideas, we develop a tool called \toolname for automatically repairing network programs from unit tests. To evaluate \toolname, we adapt $10$ benchmarks from real-world faulty network programs in Floodlight and apply \toolname to repair the benchmarks automatically. The experimental results show that \toolname is able to repair the faulty programs up to 738 lines of code for 8 benchmarks using 2 to 3 test cases, which outperforms a state-of-the-art repair tool for general Java programs. Furthermore, \toolname is very efficient in terms of repair time, and the average running time is 744 seconds across all benchmarks.

\paragraph{Contributions.} We make the following main contributions in this paper:
\begin{itemize}
    \item We present an automated program repair technique that aims to help developers debug and fix network programs automatically.
    \item We describe a fault localization approach based on symbolic execution and constraint solving for programs with imperative object-oriented features such as virtual function calls.
    \item We identify common data structures used in network programs and develop models to simplify symbolic reasoning and facilitate scalable repair.
    \item We develop a tool called \toolname based on the proposed techniques and evaluate it using $10$ benchmarks adapted from real-world network programs. The evaluation results demonstrate that \toolname is effective for fault localization and able to generate correct patches for realistic network programs.
\end{itemize}

\section{Overview} \label{sec:overview}

\begin{figure}[t]
\begin{lstlisting}[language=java]
@network
public class MacAddress {
  private long value;
  private MacAddress(long value) { this.value = value; }
  public static MacAddress NONE = new MacAddress(0);
  public static MacAddress of(long value) { return new MacAddress(value); }
  ...
}
public class FirewallRule {
  public MacAddress dl_dst;
  public boolean any_dl_dst;
  public FirewallRule() { dl_dst = MacAddress.NONE; any_dl_dst = true; ... }
  public boolean isSameAs(FirewallRule r) {
    if (... || any_dl_dst != r.any_dl_dst
            || (any_dl_dst == false &&
                dl_dst != r.dl_dst)) {
        return false;
    }
    return true;
  }
  ...
}
\end{lstlisting}
\vspace{-10pt}
\caption{Code snippet about a bug in Floodlight.}
\label{fig:example}
\end{figure}

\begin{figure}[t]
\begin{lstlisting}[language=java]
public boolean test(long mac1, long mac2) {
  FirewallRule r1 = new FirewallRule();
  r1.dl_dst = MacAddress.of(mac1); r1.any_dl_dst = false;
  FirewallRule r2 = new FirewallRule();
  r2.dl_dst = MacAddress.of(mac2); r2.any_dl_dst = false;
  boolean output = r1.isSameAs(r2);
  return output;
}
\end{lstlisting}
\vspace{-10pt}
\caption{Unit test that reveals the bug in FirewallRule.}
\label{fig:test}
\end{figure}

In this section, we give a high-level overview of our repair techniques and walk through the \toolname tool using an example adapted from the Floodlight SDN controller~\cite{floodlight-web21}.

Figure~\ref{fig:example} shows a simplified code snippet about firewall rules in Floodlight. Specifically, the program consists of two classes -- FirewallRule and MacAddress. The FirewallRule class describes rules enforced by the firewall, including information about source and destination mac addresses. The MacAddress class is an auxiliary data structure that stores the raw value of mac addresses\footnote{A unique 48-bit number that identifies each network device.} and provides functions useful to network applications.

The network program shown in Figure~\ref{fig:example} is problematic because the \texttt{isSameAs} function compares two mac addresses using the \texttt{!=} operator rather than a negation of the \texttt{equals} functions. The \texttt{!=} operator only compares two objects based on their memory addresses, whereas the intent of the developer is to check if two mac addresses have the same raw value.
The bug is revealed by the unit test shown in Figure~\ref{fig:test}, then confirmed and fixed by the Floodlight developers
\footnote{\url{https://github.com/floodlight/floodlight/commit/4d528e4bf5f02c59347bb9c0beb1b875ba2c821e}}.
In the remainder, let us illustrate how \toolname localizes this bug based on unit tests \texttt{test(1, 2) = false} and \texttt{test(1, 1) = true} and automatically synthesizes a patch to fix it.

\vspace{5pt}
\noindent \textbf{Domain-Specific Abstraction.}
\toolname has incorporated abstractions for common network data structures. For example, the MacAddress Class marked with the @network annotation contains a 48-bit integer field and several handy functions for bit manipulations. We have pre-built the high-level specifications and function summaries for data structures like MacAddress based on domain knowledge. \toolname is able to leverage the specifications and summaries for symbolic analysis without additional user input.

At a high level, \toolname enters a loop that iteratively attempts to find the fault location and synthesize the patch. Since our repair technique works in a modular fashion, \toolname first selects a function $\func$ in the program and tries to repair each possible fault location at a time. If \toolname cannot synthesize a patch consistent with the provided unit tests for any potential fault location in $\func$, it backtracks and selects the next function and repeats the same process until all possible functions are checked. We now describe the experience of running \toolname on our illustrative example.

\vspace{5pt}
\noindent \textbf{Iteration 1.}
\toolname selects the constructor of FirewallRule as the target function. Fault localization determines that the fault is located at the \texttt{dl\_dst = MacAddress.NONE} part of Line 12, because it is related to the equality checking in the unit test. However, it is not the fault location. Although \toolname invokes its underlying synthesizer and tries to synthesize a patch for this location, the synthesis procedure cannot find a solution that passes the unit test by replacing the \texttt{dl\_dst = MacAddress.NONE} statement.

\vspace{5pt}
\noindent \textbf{Iteration 2.}
\toolname selects the same function -- constructor of FirewallRule, but the fault localization switches to a different statement \texttt{any\_dl\_dst = true} at Line 12. Similar to Iteration 1, the synthesizer cannot generate a correct patch by replacing this statement.

\vspace{5pt}
\noindent \textbf{Iteration 3.}
Since none of the statements in the constructor is the fault location, \toolname now selects a different function: \texttt{isSameAs}. The fault localization determines that \texttt{any\_dl\_dst = false} at Line 15 may be the fault location as it may affect the testing results. However, having tried to replace the statement with many other candidate statements, e.g., \texttt{r.any\_dl\_dst = false}, \texttt{any\_dl\_dst = true}, the synthesizer still fails to generate the correct patch.

\vspace{5pt}
\noindent \textbf{Last iteration.}
Finally, after several attempts to localize the fault, \toolname identifies the fault lies in \texttt{dl\_dst != r.dl\_dst} at Line 16, which is indeed the reported bug location. At this time, the synthesizer manages to generate a correct patch \texttt{!dl\_dst.equals(r.dl\_dst)}. Replacing the original condition at Line 16 with this patch results in a program that can pass all the provided test cases, so \toolname has successfully repaired the original faulty program.

\section{Preliminaries} \label{sec:prelim}

In this section, we present the language of network programs and describe a program formalism that is used in the rest of paper. We also define the program repair problem that we want to solve.

\subsection{Language of Network Programs}

\begin{figure}[!t]
\[
\begin{array}{rcl}
    \emph{Program }  \prog &::=& \class^+ \\
    \emph{Class }   \class &::=& \textbf{@network} ? ~ \textbf{class } C ~ \set{ a^+ ~ \func^+ } \\
    \emph{Function } \func &::=& \textbf{function } f(x_1, \ldots, x_n) ~ (\lineid: \stmt)^+ \\
    \emph{Statement }\stmt &::=& \assignStmt{l}{e} ~|~ \jmpStmt{e}{\lineid} ~|~ \retStmt{\imm} ~|~ \newStmt{x}{C} \\
                           &|  & \staticCallStmt{x}{C}{f}{\imm_1, \ldots, \imm_n}
                           ~|~   \virtualCallStmt{x}{y}{f}{\imm_1, \ldots, \imm_n} \\
    \emph{LValue }    l    &::=& x ~|~ x.a ~|~ x[\imm] \\
    \emph{Immediate } \imm &::=& x ~|~ c \\
    \emph{Expression } e   &::=& l ~|~ c ~|~ op(e_1, \ldots, e_n) \\
\end{array}
\]
\[
\begin{array}{c}
x, y \in \textbf{Variable} \quad c \in \textbf{Constant} \quad \lineid \in \textbf{LineID} \\
C \in \textbf{ClassName} \quad f, f_0 \in \textbf{FuncName} \quad a \in \textbf{FieldName} \\
\end{array}
\]
\caption{Syntax of network programs.}
\label{fig:syntax}
\vspace{-10pt}
\end{figure}

The language of network programs considered in this paper is summarized in Figure~\ref{fig:syntax}. A network program consists of a set of classes, where each class has an optional annotation \textrm{@network} to denote that the class is a network-related data structure such as mac address and IPv4 address. The annotations are helpful for recognizing network data structures and facilitate domain-specific analysis during program repair.

Each class in the program consists of a list of fields and functions. Each function has a name, a parameter list, and a function body. We collectively refer to the function name and parameter list as the \emph{function signature}. The function body is a list of statements, where each statement is labeled with its line number.
Various kinds of statements are included in our language of network programs. Specifically, assign statement $\assignStmt{l}{e}$ assigns expression $e$ to left value $l$. Conditional jump statement $\jmpStmt{e}{L}$ first evaluates predicate $e$. If the result is true, then the control flow jumps to line $L$; otherwise, it performs no operation. Note that our language does not have traditional if statements or loop statements, but those statements can be expressed using conditional jumps.~\footnote{Our repair techniques only handle bounded loops. If there are unbounded loops in the network program, we need to perform loop unrolling.
}
Return statement $\retStmt{v}$ exits the current function with return value $v$. New statement $\newStmt{x}{C}$ creates an object of class $C$ and assigns the object address to variable $x$. Static call $\staticCallStmt{x}{C}{f}{v_1, \ldots, v_n}$ invokes the static function $f$ in class $C$ with arguments $v_1, \ldots, v_n$ and assigns the return value to variable $x$. Similarly, virtual call $\virtualCallStmt{x}{y}{f}{v_1, \ldots, v_n}$ invokes the virtual function $f$ on \emph{receiver} object $y$ with arguments $v_1, \ldots, v_n$ and assigns the return value to variable $x$.
Different kinds of expressions are supported including constants, variable accesses, field accesses, array accesses, arithmetic operations, and logical operations.
Since the semantics of network programs is similar to that of traditional programs written in object-oriented languages, we omit the formal description of semantics.




As is standard, we assume class names are implicitly appended to function names, so a  function signature uniquely determines a function in the program. In addition, we assume each statement in the program is labeled with a globally unique line number, and line numbers are consecutive within a function.
In the remainder of this paper, we use several auxiliary functions and relations about the control flow structure of programs. Specifically, $\textsf{FirstLine}(\prog, \func)$ returns the first line of function $\func$ in program $\prog$. $\textsf{IsPrevLine}(\prog, \lineid_1, \lineid_2)$ is a relation representing that $\lineid_1$ is a control-flow predecessor of $\lineid_2$ in program $\prog$: $\lineid_1$ is a predecessor of $\lineid_2$ if (1) $\lineid_2$ is the next line of $\lineid_1$ and $\lineid_1$ is not a return statement, or (2) $\lineid_1$ is a jump statement and the target location of jump is $\lineid_2$.


\subsection{Problem Statement}

In this paper, we assume a unit test $t$ is written in the form of a pair $(I, O)$, where $I$ is the input and $O$ is the expected output. Given a network program $\prog$ and a unit test $t = (I, O)$, we say $\prog$ passes the test $t$ if executing $\prog$ on input $I$ yields the expected output $O$, denoted by $\denot{\prog}_I = O$. Otherwise, if $\denot{\prog}_I \neq O$, we say $\prog$ fails the test $t$.
In general, given a network program $\prog$ and a set of unit tests $\exs$, program $\prog$ is \emph{faulty} modulo $\exs$ if there exists a test $t \in \exs$ such that $\prog$ fails on $t$.

Now let us turn the attention to the meaning of fault locations and patches.

\begin{definition}[Fault location and patch]
Let $\prog$ be a program that is faulty modulo tests $\exs$. Line $\lineid$ is called the \emph{fault location} of $\prog$, if there exists a statement $s$ such that replacing line $\lineid$ of $\prog$ with $s$ yields a new program that can pass all tests in $\exs$. Here, the statement $s$ is called a \emph{patch} to $\prog$.
\end{definition}

Having defined these concepts, we precisely describe our research problem in the next.

\paragraph{\textbf{Problem statement.}}
Given a network program $\prog$ that is faulty modulo tests $\exs$, our goal is to find a fault location $\lineid$ in $\prog$ and generate the corresponding patch $s$, such that for any unit test $t \in \exs$, the patched program $\prog'$ can always pass the test $t$.

\section{Modular Program Repair} \label{sec:repair}

In this section, we present our algorithm for automatically repairing network programs from a set of unit tests.

\subsection{Algorithm Overview} \label{sec:algo-overview}

\begin{figure}[!t]
\begin{algorithm}[H]
\caption{Modular Program Repair}
\label{algo:repair}
\begin{algorithmic}[1]
\Procedure{\textsc{Repair}}{$\prog, \exs$}
\vspace{2pt}
\small
\Statex \textbf{Input:} Program $\prog$, examples $\exs$
\Statex \textbf{Output:} Repaired program $\prog'$ or $\bot$ to indicate failure
\State $\prog \gets \textsf{Abstraction}(\prog)$;
\State $\visited \gets \set{\lineid \mapsto \emph{false} ~|~ \lineid \in \textsf{Lines}(\prog)}$; $\prog' \gets \bot$;
\While{$\prog' = \bot$}
    \State $\func \gets \textsf{SelectFunction}(\prog, \visited)$;
    \If{$\func = \bot$} \Return $\bot$; \EndIf
    \State $\visited, \prog' \gets \textsc{RepairFunction}(\prog, \func, \exs, \visited)$;
\EndWhile
\State \Return $\prog'$;
\EndProcedure
\vspace{5pt}

\Procedure{\textsc{RepairFunction}}{$\prog, \func, \exs, \visited$}
\Statex \textbf{Input:} Program $\prog$, function $\func$, examples $\exs$, visited map $\visited$
\Statex \textbf{Output:} Updated visited map $\visited$, repaired program $\prog'$
\State $\prog' \gets \bot$;
\While{$\prog' = \bot$}
    \State $\lineid \gets \textsc{LocalizeFault}(\prog, \func, \exs, \visited)$;
    \If{$\lineid \neq \bot$}
        \State $\visited \gets \visited[\lineid \mapsto \emph{true}]$;
    \Else
        \State $\visited \gets \visited[\lineid' \mapsto \emph{true} ~|~ \textsf{TransInFunc}(\lineid', \prog, \func)]$;
    \EndIf
    \If {$\lineid = \bot$ \textbf{or} $\textsf{IsCallStmt}(\prog, \lineid)$} \Return $\visited, \bot$; \EndIf
    \State $\prog' \gets \textsc{SynthesizePatch}(\prog, \exs, \func, \lineid)$;
\EndWhile
\State \Return $\visited, \prog'$;
\EndProcedure
\end{algorithmic}
\end{algorithm}
\vspace{-10pt}
\end{figure}

The top-level repair algorithm is described in Algorithm~\ref{algo:repair}. The \textsc{Repair} procedure takes as input a faulty network program $\prog$ and unit tests $\exs$ and produces as output a repaired program $\prog'$ or $\bot$ to indicate repair failure.

At a high level, the \textsc{Repair} procedure maintains a visited map $\visited$ from line numbers to boolean values, representing whether each line of $\prog$ is checked or not. Specifically, $\visited[\lineid] = \emph{true}$ indicates line $\lineid$ is checked; otherwise, $\visited[\lineid] = \emph{false}$ means $\lineid$ is not checked yet. As shown in Algorithm~\ref{algo:repair}, the \textsc{Repair} procedure first applies the domain-specific abstraction to program $\prog$ (Line 2) and initializes the visited map $\visited$ by setting every line in $\prog$ as not checked (Line 3).
Next, it tries to iteratively repair $\prog$ in a modular way until it finds a program $\prog'$ that is not faulty modulo tests $\exs$ (Lines 4 -- 8). In particular, the \textsc{Repair} procedure invokes \textsf{SelectFunction} to choose a function $\func$ as the target of repair (Line 5). If none of the functions in $\prog$ can be repaired, it returns $\bot$ to indicate that the repair procedure failed (Line 6). Otherwise, it invokes the \textsc{RepairFunction} procedure (Line 7) to enter the localization-synthesis loop inside the target function $\func$.

Focusing on one function at a time and repairing programs in a modular fashion is beneficial for two reasons. First, given a faulty program $\prog$ and a target function $\func$, we only need to check functions  in the call stack from the main function of $\prog$ to $\func$. It can significantly reduce the number of functions to analyze and the number of locations to check, which enables faster program analysis and repair. Second, we can summarize the behavior of non-target functions and reuse summaries to achieve better scalability. Specifically, given a target function $\func$, all functions that are not in the call stack of $\func$ can be replaced with their corresponding summaries. It further decreases the number of locations to track, because we do not need to inline non-target functions that are invoked in the transitive callers of $\func$.

In addition to the program $\prog$ and tests $\exs$, the \textsc{RepairFunction} procedure takes as input a target function $\func$ and the current visited map $\visited$. It produces as output the updated version of the visited map $\visited$, as well as a repaired program $\prog'$ or $\bot$ to indicate that the function $\func$ cannot be repaired. As shown in Lines 11 -- 18 of Algorithm~\ref{algo:repair}, \textsc{RepairFunction} alternatively invokes sub-procedures \textsc{LocalizeFault} and \textsc{SynthesizePatch} to repair the target function.
In particular, the goal of \textsc{LocalizeFault} is to identify a fault location in function $\func$. If \textsc{LocalizeFault} manages to find a fault location $\lineid$ in $\func$, then line $\lineid$ is marked as visited (Line 14). Otherwise, if \textsc{LocalizeFault} returns $\bot$, it means function $\func$ and all functions transitively invoked in $\func$ are correct or not repairable. In this case, all lines in $\func$ and its transitive callees are marked as checked (Line 16).
Furthermore, if the identified fault location $\lineid$ corresponds to a statement that invokes $\func'$, it means the fault location is inside $\func'$. Thus, \textsc{RepairFunction} directly returns $\bot$ (Line 17) and \textsf{SelectFunction} will choose $\func'$ as the target function in the next iteration. 
On the other hand, the goal of the sub-procedure \textsc{SynthesizePatch} is generating a patch for function $\func$ given the fault location $\lineid$. If \textsc{SynthesizePatch} successfully synthesizes a patch and produces a non-faulty program $\prog'$, then the entire procedure succeeds with repaired program $\prog'$. Otherwise, \textsc{RepairFunction} backtracks with a new program location and repeat the same process.

In the rest of this section, we explain domain-specific abstraction, fault localization, and patch synthesis in more details.

\subsection{Domain-Specific Abstraction} \label{sec:abstraction}

Given a network program $\prog$, we identify all classes with the \textrm{@network} annotation in $\prog$ and replace their functions
\footnote{We assume getter and setter functions are implicitly defined for public fields. Load and store operations on public fields are performed through getter and setter functions.}
with corresponding abstractions based on domain knowledge. The abstraction $\abs[\func]$ of a function $\func$ is an over-approximation of $\func$ that is precise enough to characterize the behavior of $\func$.
Our insight is that using domain specific abstractions for network data structures can significantly help the program analysis and repair process. This insight is based on two key observations.

First, source code for network programs may only be partially available due to the use of high-level interface and native implementation. For example, while the interface of an operation is defined in Java, its implementation may only be available in C or C++. Although it is possible to analyze programs written in multiple languages, such analysis imposes unnecessary overhead because it typically requires different infrastructures and techniques for different languages. Using domain specific abstractions allows us to directly utilize the high-level specifications of network operations, and thus provides a unified way for us to reason about the program behavior.

\begin{example}
Consider the \textrm{equals} function of a mac address:
\begin{lstlisting}[language=java, numbers=none]
@network public class MacAddress {
  private long value; ...
  public boolean equals(final Object obj) {
    if (this == obj) return true;
    if (obj == null) return false;
    if (getClass() != obj.getClass()) return false;
    if (value != ((MacAddress) obj).value) return false;
    return true;
  }
}
\end{lstlisting}
Here, the implementation of the function \textrm{getClass} is not available. Since we know the \textrm{equals} function essentially checks whether two mac addresses are identical or not, we can use the following high-level abstraction
\[
\abs[\textrm{equals}]: \lambda x.~ \lambda y.\,(x.\emph{dtype} = y.\emph{dtype} \land x.\emph{value} = y.\emph{value}\,),
\]
where $x.\emph{dtype}$ denotes the dynamic type of the object $x$.
\end{example}

Second, network programs have complex operations that are challenging for symbolic reasoning. For instance, bit manipulations are heavily used in network data structures. While bit manipulations can improve the performance of network programs, they present significant challenges for symbolic analysis due to the encoding in the  theory of bitvectors. Leveraging domain knowledge in network systems, we can define abstractions that over-approximates functions involving bit manipulations but use simpler encoding that is more amenable to symbolic reasoning.

\begin{example}
Consider the \textrm{hashCode} function of a mac address that involves bit manipulations
\begin{lstlisting}[language=java, numbers=none]
@network public class MacAddress {
  private long value; ...
  public int hashCode() {
    final int prime = 31; int result = 1;
    result = prime * result + (int) (value ^ (value >>> 32));
    return result;
  }
}
\end{lstlisting}
The \textrm{hashCode} function only needs to ensure that if two mac addresses $x$ and $y$ are identical, then $\textrm{hashCode}(x) = \textrm{hashCode}(y)$. Based on this contract, we can simply define its abstraction as
\[
\abs[\textrm{hashCode}]: \lambda x.~ x.value.
\]
This abstraction does not involve any bit manipulation but it is precise enough for subsequent symbolic analysis.
\end{example}

\subsection{Fault Localization} \label{sec:localization}

Next, we present our fault localization technique that aims to find the fault location in a given target function. Before delving into the details of the algorithm, we will explain the methodology of fault localization at a high level.

\subsubsection{Methodology} \label{sec:localization-method} \hfill

At a high level, our fault localization technique uses a symbolic approach by reducing the fault localization problem into a constraint solving problem. In particular, we introduce a boolean variable for each line $\lineid$, denoted by $\lineMap[\lineid]$, and encode the fault localization problem as an SMT formula, such that the value of the variable $\lineMap[\lineid]$ indicates whether line $\lineid$ is correct or not.

\vspace{5pt}
\noindent \textbf{Checking faulty programs.}
To understand how to encode the fault localization problem, let us first explain how to encode the consistency check given a program $\prog$ and a test case $t = (I, O)$. Specifically, the encoded SMT formula $\cstr(t)$ consists of three components:
\begin{enumerate}
\item \emph{Semantic constraints}. For each line $\lineid_i: \stmt_i$, we generate a formula $\cstr_i(S, S')$ to describe the semantics of the statement $\stmt_i$. Specifically, given a state $S$ that holds before statement $\stmt_i$, $\cstr_i(S, S')$ is valid if $S'$ is the state after executing $\stmt_i$.
\item \emph{Control flow integrity constraints}. In order to ensure all traces satisfying the constraint faithfully follow the control flow structure of a given program $\prog$, we generate constraints $\cstr_f$ to describe the order of statements in $\prog$. For example, if there are statements $\stmt_1; \stmt_2; \stmt_3$ in $\prog$, formula $\cstr_f$ ensures a trace only considers order $\stmt_1, \stmt_2, \stmt_3$, but not $\stmt_1, \stmt_3, \stmt_2$.
\item \emph{Consistency between program and test}. For the provided test case $t = (I, O)$, we also generate formula $\cstr_{in}(S_0, I)$ and $\cstr_{out}(S_n, O)$ to ensure the program behavior is consistent with the test. In particular, $\cstr_{in}(S_0, I)$ binds input $I$ to the initial state $S_0$ and $\cstr_{out}(S_n, O)$ describes the connection between output $O$ and final state $S_n$.
\end{enumerate}

The satisfiability of formula $\cstr(t)$ indicates the result of consistency check. If $\cstr(t)$ is satisfiable, then program $\prog$ can pass the test $t$, because there exists a valid trace according to the control flow and every pair of adjacent states in this trace is consistent with the semantics of the corresponding statement. Otherwise, if formula $\cstr(t)$ is unsatisfiable, then $\prog$ fails the test $t$.

Now to check whether program $\prog$ is faulty modulo a set of unit tests $\exs$, we can conjoin the formula $\cstr(t_j)$ for each unit test $t_j \in \exs$ and obtain the conjunction
\[
\cstr = \bigwedge_{t_j \in \exs} \cstr(t_j)
\]
Here, the satisfiability of formula $\cstr$ indicates whether $\prog$ is faulty modulo tests $\exs$.

\vspace{5pt}
\noindent \textbf{Methodology of fault localization.}
Let $\prog$ be a faulty program modulo $\exs$, we know the corresponding formula $\cstr$ for consistency check is unsatisfiable. Suppose the fault location is line $\lineid_i$, one key insight is that replacing the semantic constraint $\cstr_i(S, S')$ with $\emph{true}$ yields a satisfiable formula. This is because $\emph{true}$ does not enforce any constraint between the pre-state $S$ and post-state $S'$, so a previously invalid trace caused by the bug at $\lineid$ becomes valid now.

Based on this insight, we develop a methodology to find the fault location using symbolic reasoning. Specifically, given a consistency check formula $\cstr$, we can obtain a fault localization formula $\cstr'$ by replacing the semantic constraint $\cstr_i(S, S')$ with $\lineMap[\lineid_i] \to \cstr_i(S, S')$ for every line $\lineid_i, i \in [1, n]$. Here, variable $\lineMap[\lineid_i]$ decides whether or not it turns the semantic constraint of $\lineid_i$ into $\emph{true}$. Thus, $\lineMap[\lineid_i] = \emph{false}$ indicates $\lineid_i$ is a fault location.

One hiccup here is that formula $\cstr'$ is always satisfiable and a model of $\cstr'$ can simply assign $\lineMap[\lineid_i] = \emph{false}$ for all $\lineid_i$. It means all lines in the program are fault locations, which is not useful for fault localization. To address this issue, we can add a cardinality constraint stating there are exactly $K$ variables in map $\lineMap$ that can be assigned to \emph{false}, which forces the constraint solver to find exactly $K$ fault locations in program $\prog$.

\subsubsection{Algorithm} \hfill

\begin{figure}[!t]
\vspace{-10pt}
\begin{algorithm}[H]
\caption{Fault Localization}
\label{algo:localize}
\begin{algorithmic}[1]
\Procedure{\textsc{LocalizeFault}}{$\prog, \func, \exs, \visited$}
\vspace{2pt}
\small
\Statex \textbf{Input:} Program $\prog$, function $\func$, examples $\exs$, visited map $\visited$
\Statex \textbf{Output:} Buggy line $\lineid$ or $\bot$ to indicate failure

\State $\lineMap \gets \set{}$; $\pathMap \gets \set{}$; $\memMap \gets \set{}$;
\For{$\lineid \in \dom(\visited)$}
    \If{$\visited[\lineid] = \emph{true}$}
        \State $\lineMap[\lineid] \gets \emph{true}$;
    \EndIf
\EndFor
\State $\summary \gets \set{ \func' \mapsto \textsf{Summary}(\func') ~|~ \func' \in \textsf{GetCallees}(\prog, \func)}$;
\State $\cstr \gets \textsc{Encode}(\lineMap, \summary, \pathMap, \memMap, \func)$;
\State $\cstr \gets \cstr \land \textsf{ExampleConsistency}(\prog, \exs)$;
\If{\textsf{UNSAT}($\cstr$)} \Return $\bot$; \EndIf
\State \Return $\textsf{Filter}(\textsf{IsFalse}, \textsf{GetModel}(\cstr))$;
\EndProcedure
\end{algorithmic}
\end{algorithm}
\vspace{-10pt}
\end{figure}

Having explained the high-level methodology, let us look at the detailed fault localization algorithm for network programs.
As shown in Algorithm~\ref{algo:localize}, the \textsc{LocalizeFault} procedure takes as input a program $\prog$, a target function $\func$, a set of tests $\exs$, and a visited map $\visited$, and produces as output the fault location in $\func$ that causes the behavior of $\prog$ is not consistent with $\exs$.

In the beginning, the \textsc{LocalizeFault} procedure initializes the boolean map $\lineMap$ from visited map $\visited$ (Lines 2 -- 3). If line $\lineid$ is marked as visited by $\visited$, then $\lineMap[\lineid]$ is initialized to $\emph{true}$ because $\lineid$ is not a fault location. Otherwise, $\lineMap[\lineid]$ does not have a determined value. The initialization also creates two empty maps $\pathMap$ and $\memMap$.
Here, $\pathMap$ is a mapping related to the encoding control flow integrity. It maps a line number $\lineid$ to a boolean variable $\pathMap[\lineid]$, where $\pathMap[\lineid] = \emph{true}$ represents line $\lineid$ occurs in the trace.
$\memMap$ maps a line $\lineid$ to an uninterpreted function $\memMap[\lineid]$, representing the memory after executing line $\lineid$. In particular, we use an uninterpreted function to represent the memory, which takes an address $x$ as input and produces as output the value stored at address $x$. Furthermore, since we need to maintain multiple versions of the memory based on the execution status of each line in $\prog$, we introduce a map $\memMap$ from line numbers to their corresponding uninterpreted functions.

Next, \textsc{LocalizeFault} computes function summaries for all callees of target function $\func$ and follows the methodology in Section~\ref{sec:localization-method} to localize fault based on symbolic reasoning. Specifically, it invokes the \textsc{Encode} procedure to generate semantic constraints and control flow integrity constraints (Line 7) and then invokes \textsf{ExampleConsistency} to generate consistency check for provided test $\exs$ (Line 8). If the generated formula $\cstr$ is unsatisfiable, fault localization fails for target function $\func$ (Line 9). Otherwise, \textsc{LocalizeFault} returns the line $\lineid$ where the corresponding variable $\lineMap[\lineid] = \emph{false}$ based on the model of $\cstr$ (Line 10).

Since the encoding for binding test cases to initial and final states is straightforward, we omit the discussion. In the remainder, we describe how to generate semantic constraint and control flow integrity constraint in more detail.

\begin{figure}[!t]
\centering
\[
\begin{array}{c}
\irulelabel
{}
{\memMap, \lineid \vdash c \twoheadrightarrow c}
{\textrm{(Const)}} \quad

\irulelabel
{\begin{array}{c}
    \memMap, \lineid \vdash l \hookrightarrow \addr_l \quad l \in \set{x, x.a, x[v]}
\end{array}}
{\memMap, \lineid \vdash l \twoheadrightarrow \memMap[\lineid](\addr_l)}
{\textrm{(LValue)}} \\ \ \\

\irulelabel
{\memMap, \lineid \vdash e_i \twoheadrightarrow \term_{e_i} \quad i = 1, \ldots, n}
{\memMap, \lineid \vdash op(e_1, \ldots, e_n) \twoheadrightarrow op(\term_{e_1}, \ldots, \term_{e_n})}
{\textrm{(Op)}}

\end{array}
\]
\caption{Inference rules for encoding expressions.}
\label{fig:rules-expr}
\end{figure}

\begin{figure}[!t]
\centering
\[
\hspace{-10pt}
\begin{array}{c}
\irulelabel
{}
{\memMap, \lineid \vdash x \hookrightarrow \text{Addr}(x)}
{\!\!\!\!\textrm{(Var)}} \

\irulelabel
{\begin{array}{c}
    i = \text{offset}(a) \\
    \memMap, \lineid \vdash x \hookrightarrow \addr_{x} \\
\end{array}}
{\memMap, \lineid \vdash x.a \hookrightarrow \addr_{x} + i}
{\!\!\!\!\textrm{(Field)}} \

\irulelabel
{\begin{array}{c}
    \memMap, \lineid \vdash v \twoheadrightarrow i \\
    \memMap, \lineid \vdash x \hookrightarrow \addr_{x} \\
\end{array}}
{\memMap, \lineid \vdash x[v] \hookrightarrow \addr_{x} + i}
{\!\!\!\!\textrm{(Array)}}

\end{array}
\]
\caption{Inference rules for encoding the address of left-values.}
\label{fig:rules-addr}
\end{figure}

\vspace{5pt}
\noindent \textbf{Expressions.}
Since the semantic constraint of a line involves encoding expressions, we first present the symbolic encoding of expressions in our network programs.
The inference rules of generating constraints for expressions are summarized in Figure~\ref{fig:rules-expr}. A judgment of the form
\[
\memMap, \lineid \vdash e \twoheadrightarrow \phi
\]
denotes that the encoding of expression $e$ is $\phi$ given memory map $\memMap$ and line number $\lineid$. For example, the \textrm{Const} rule states that constant $c$ is encoded as $c$. To encode a left-value $l$, including variable, field, and array accesses, we first need to obtain the address $\addr_l$ of $l$. Then according to the \textrm{LValue} rule, we look up the memory map $\memMap$ based on the current line number $\lineid$ and address $\addr_l$ to get its value $\memMap[\lineid](\addr_l)$. For an expression $op(e_1, \ldots, e_n)$, we can recursively encode sub-expression $e_i$ as $\psi_{e_i}$ and generate the composed encoding $op(\psi_{e_1}, \ldots, \psi_{e_n})$ (rule \textrm{Op}).

Similarly, inference rules of encoding addresses are summarized in Figure~\ref{fig:rules-addr}, where judgements of the form
\[
\memMap, \lineid \vdash e \hookrightarrow \phi
\]
denote the address of expression $e$ is $\phi$. Specifically, the address of variable $x$ is simply obtained by the address operator (rule \textrm{Var}). The address of field access $x.a$ is $\addr_x + i$ where $\addr_x$ is the address of $x$ and $i$ is the offset of field $a$. Similarly, the address of array access $a[v]$ is $\addr_x + i$ where $\addr_x$ is the address of $x$ and $i$ is the symbolic encoding of immediate number $v$.

\begin{figure}[!t]
\centering
\[
\hspace{-10pt}
\begin{array}{c}
\irulelabel
{\begin{array}{c}
    \memMap, \lineid \vdash e \twoheadrightarrow \term_e \quad
    \memMap, \lineid \vdash l \hookrightarrow \addr_l \\
    \lineid' = \textrm{PrevLine}(\lineid, \pathMap, \func) \quad
    \cstr_m \equiv \lineMap[\lineid] \to \memMap[\lineid](\addr_l) = \term_e \\
    \cstr_c \equiv \bigwedge_{z \neq \addr_l} \memMap[\lineid](z) = \memMap[\lineid'](z) \quad
    \cstr_f \equiv \pathMap[\lineid+1] \land \pathMap[\lineid'] \\
\end{array}}
{\lineMap, \summary, \pathMap, \memMap, \func \vdash \lineid: \assignStmt{l}{e} \leadsto
\pathMap[\lineid] \to \cstr_m \land \cstr_c \land \cstr_f }
{\textrm{(Assign)}} \\ \ \\

\irulelabel
{\begin{array}{c}
    \memMap, \lineid \vdash e \twoheadrightarrow \term_e \quad
    \lineid' = \textrm{PrevLine}(\lineid, \pathMap, \func) \\ 
    \cstr_c \equiv \bigwedge_{z} \memMap[\lineid](z) = \memMap[\lineid'](z) \quad
    \cstr_f \equiv (\pathMap[\lineid''] \lor \pathMap[\lineid+1]) \land \pathMap[\lineid'] \\
    \cstr_m \equiv ((\lineMap[\lineid] \to \term_e) \leftrightarrow \pathMap[\lineid'']) \land ((\lineMap[\lineid] \to \neg \term_e) \leftrightarrow \pathMap[\lineid+1])
\end{array}}
{\lineMap, \summary, \pathMap, \memMap, \func \vdash \lineid: \jmpStmt{e}{\lineid''} \leadsto
\pathMap[\lineid] \to \cstr_m \land \cstr_c \land \cstr_f }
{\textrm{(Jump)}} \\ \ \\

\irulelabel
{\begin{array}{c}
    \memMap, \lineid \vdash x \twoheadrightarrow \term_x \quad
    \cstr_f \equiv \pathMap[\lineid+1] \land \pathMap[\lineid'] \\
    \cstr_c \equiv \bigwedge_{z} \memMap[\lineid](z) = \memMap[\lineid'](z) \land \emph{DType}[\term_x] = C \\
\end{array}}
{\lineMap, \summary, \pathMap, \memMap, \func \vdash \lineid: \newStmt{x}{C} \leadsto
\pathMap[\lineid] \to \cstr_c \land \cstr_f }
{\textrm{(New)}} \\ \ \\

\irulelabel
{\begin{array}{c}
    \memMap, \lineid \vdash v \twoheadrightarrow \term_v \quad
    \lineid' = \textrm{PrevLine}(\lineid, \pathMap, \func) \\ 
    \cstr_m \equiv \lineMap[\lineid] \to \memMap[\lineid](\addr_{ret}) = \term_v \quad
    \cstr_c \equiv \bigwedge_{z \neq \addr_{ret}} \memMap[\lineid](z) = \memMap[\lineid'](z) \quad
    \cstr_f \equiv \pathMap[\lineid'] \\
\end{array}}
{\lineMap, \summary, \pathMap, \memMap, \func \vdash \lineid: \retStmt{v} \leadsto 
\pathMap[\lineid] \to \cstr_m \land \cstr_c \land \cstr_f }
{\textrm{(Return)}} \\ \ \\

\irulelabel
{\begin{array}{c}
    \memMap, \lineid \vdash v_i \twoheadrightarrow \term_i \quad i = 1, \ldots, n \quad
    \lineid' = \textrm{PrevLine}(\lineid, \pathMap, \func) \\
    \cstr_1 = \summary[C.f][\term_1/\emph{arg}_1, \ldots, \term_n/\emph{arg}_n, x/\emph{ret}, \memMap[\lineid']/\memMap_{in}, \memMap[\lineid]/\memMap_{out}] \\
    \cstr_m \equiv \lineMap[\lineid] \to \cstr_1 \quad
    \cstr_f \equiv \pathMap[\lineid+1] \land \pathMap[\lineid'] \\
\end{array}}
{\lineMap, \summary, \pathMap, \memMap, \func \vdash \lineid: \staticCallStmt{x}{C}{f}{v_1,\ldots,v_n} \leadsto
\pathMap[\lineid] \to \cstr_m \land \cstr_f
}
{\textrm{(SCall)}} \\ \ \\

\irulelabel
{\begin{array}{c}
    \memMap, \lineid \vdash v_i \twoheadrightarrow \term_i \quad i = 1, \ldots, n \quad
    \memMap, \lineid \vdash y \twoheadrightarrow \term_y \quad  
    \lineid' = \textrm{PrevLine}(\lineid, \pathMap, \func) \\
    \cstr_{S_j} \equiv \summary[C_j.f][\term_y/\emph{this},\term_1/\emph{arg}_1, \ldots, \term_n/\emph{arg}_n, x/\emph{ret}, \memMap[\lineid']/\memMap_{in}, \memMap[\lineid]/\memMap_{out}] \\
    \cstr_1 \equiv \bigwedge_{j=1,\ldots,m} \emph{DType}[\term_y] = C_j \to \cstr_{S_j} \quad
    C_j <: \text{Class}(f) \quad j = 1, \ldots, m \\
    \cstr_m \equiv \lineMap[\lineid] \to \cstr_1 \quad
    \cstr_f \equiv \pathMap[\lineid+1] \land \pathMap[\lineid'] \\
\end{array}}
{\lineMap, \summary, \pathMap, \memMap, \func \vdash \lineid: \virtualCallStmt{x}{y}{f}{v_1,\ldots,v_n} \leadsto
\pathMap[\lineid] \to \cstr_m \land \cstr_f
}
{\textrm{(VCall)}} \\ \ \\

\irulelabel
{\begin{array}{c}
    \lineMap, \summary, \pathMap, \memMap, \func \vdash \lineid_i: \stmt_i \leadsto \cstr_i \quad i=1,\ldots,n \\
\end{array}}
{\lineMap, \summary, \pathMap, \memMap, \func \vdash \lineid_1: \stmt_1; \ldots; \lineid_n: \stmt_n \leadsto \cstr_1 \land \ldots \land \cstr_n}
{\textrm{(Compose)}}

\end{array}
\]
\caption{Inference rules for generating semantic constraints and control flow integrity constraints. Given a function $f$ with $n$ arguments, the formal parameters of $f$ are denoted by $\emph{arg}_1, \ldots, \emph{arg}_n$. The return variable is denoted by $\emph{ret}$, and the reference variable is denoted by $\emph{this}$.}
\label{fig:rules-localize}
\end{figure}

\vspace{5pt}
\noindent \textbf{Statements.}
Having explained the constraint generation for expressions, now let us illustrate how to generate constraints for statements. As shown in Figure~\ref{fig:rules-localize}, our inference rules for statement-level constraints take judgments of the form
\[
\lineMap, \summary, \pathMap, \memMap, \func \vdash \lineid: \stmt \leadsto \cstr
\]
meaning that the statement $\stmt$ at line $\lineid$ is encoded as formula $\cstr$ under line indicator map $\lineMap$, summary map $\summary$, trace selector map $\pathMap$, memory map $\memMap$, and target function $\func$.
For ease of illustration, we divide the final constraints into three parts: $\cstr_m$ denotes the semantic constraint about the ``maybe incorrect'' operations in the statement, which typically involves updating the memory. $\cstr_c$ represents the semantic constraint about the ``always correct'' operations in the statement, which usually describes what memory values should remain unchanged by the execution of the statement. $\cstr_f$ is the control flow integrity constraint that characterizes a valid trace based on the control flow structure of function $\func$.

\vspace{5pt}
\noindent \textbf{Assign statement.}
Given an assign statement $\assignStmt{l}{e}$ at line $\lineid$, the \textrm{Assign} rule generates a formula $\pathMap[\lineid] \to \cstr_m \land \cstr_c \land \cstr_f$, which means if line $\lineid$ is selected in the trace, then all three kinds of constraints should hold.
Specifically, it first computes the address of left-hand side $l$ as $\addr_l$ and computes the expression encoding of right-hand side $e$ as $\psi_e$. The generated ``maybe incorrect'' constraint adds a guard $\lineMap[\lineid]$ to memory update $\memMap[\func][a]$, saying if the line is not the fault location, then the value at address $\addr_l$ after executing line $\lineid$ is $\psi_e$. However, if line $\lineid$ is the fault location, then no constraint is added effectively. The ``always correct'' constraint states all values except the one at $\addr_l$ should be preserved by the assign statement. The control flow integrity constraint $\cstr_f$ says that both previous line $\lineid'$ and next line $\lineid+1$ must be selected in the trace.

\vspace{5pt}
\noindent \textbf{Jump statement.}
Similar to assign statements, the \textrm{Jump} rule also emits three constraints if the statement at line $\lineid$ is a conditional jump statement $\jmpStmt{e}{\lineid''}$. Since the jump condition $e$ might be faulty, the rule adds a guard $\lineMap[\lineid]$ to the encoded expression $\psi_e$. $\cstr_m$ says if line $\lineid$ is not the fault location, then the jump destination $\lineid''$ is selected in the trace if condition evaluates to $\emph{true}$ or the next line $\lineid$ is selected in the trace if condition $e$ evaluates to $\emph{false}$. Furthermore, the control flow integrity constraint $\cstr_f$ requires that (1) the previous line $\lineid'$ must occur in the trace, and (2) either the next line $\lineid+1$ or the jump destination $\lineid''$ must occur in the trace. In addition, since a jump statement does not write to the memory, $\cstr_c$ describes all values in the previous memory $\memMap[\lineid']$ are preserved in current memory $\memMap[\lineid]$.

\vspace{5pt}
\noindent \textbf{New statement.}
Since we do not consider memory allocation as a source of bugs, the \textrm{New} rule does not generate the ``maybe incorrect'' constraint $\cstr_m$. Instead, given a statement $\newStmt{x}{C}$, it generates $\cstr_c$ stating the dynamic type of $x$ is $C$ and all values in the member are preserved. In addition, the previous line $\lineid'$ and next line $\lineid+1$ must occur in the trace.

\vspace{5pt}
\noindent \textbf{Return statement.}
Given a return statement $\retStmt{v}$ at line $\lineid$, the \textrm{Return} rule first evaluates immediate number $v$ to $\psi_v$, and then write the value to memory $\memMap[\lineid]$ at location $\addr_{ret}$, where $\emph{ret}$ is an implicit variable for storing return values and $\addr_{ret}$ is its address. Since the return value could be faulty, the rule adds a guard $\lineMap[\lineid]$ in constraint $\cstr_m$. By contrast, all other values are considered correct, so it preserves all but the return value after execution in constraint $\cstr_c$. In addition, the control flow integrity constraint $\cstr_f$ only requires the previous line to occur in the trace.

\vspace{5pt}
\noindent \textbf{Static call.}
Since our fault localization algorithm is modular and summaries are computed for all functions in the program, we can directly utilize the function summary for invocations. In particular, given a static function call $\staticCallStmt{x}{C}{f}{v_1, \ldots, v_n}$, the \textrm{SCall} rule evaluates the actual parameter $v_i$ to $\psi_i$ and substitutes the formal parameter $\emph{arg}_i$ in summary $\summary[C.f]$ with $\psi_i$. Furthermore, it also substitutes the return variable $\emph{ret}$ with variable $x$ and substitutes the formal input memory $\memMap_{in}$ and output memory $\memMap_{out}$ with the actual memories $\memMap[\lineid']$ and $\memMap[\lineid]$, respectively.

\vspace{5pt}
\noindent \textbf{Virtual call.}
The \textrm{VCall} rule for virtual function calls is similar to \textrm{SCall}. The only difference is that it needs to dispatch function summaries based on the receiver object. Recall that every time the program creates a new object, the \textrm{New} rule stores its dynamic type in the $\emph{DType}$ map. Thus, given a virtual call $\virtualCallStmt{x}{y}{f}{v_1, \ldots, v_n}$, \textrm{SCall} can obtain the dynamic type of receiver object $y$ by evaluating $y$ to $\psi_y$ and looking up the map $\emph{DType}$. According to the dynamic type $\emph{DType}[\psi_y]$, \textrm{SCall} selects the appropriate function summary to use.

\vspace{5pt}
\noindent \textbf{Statement composition.}
Finally, the \textrm{Compose} rule is quite straightforward. Specifically, the constraints generated for multiple statements are obtained inductively by conjoining the constraints for each individual statement.


\subsection{Patch Synthesis} \label{sec:synthesis}

The last step of our repair algorithm is to generate a patch to fix the faulty program. The high-level idea is to reduce the patch generation problem to an expression synthesis problem. Specifically, given a faulty function $\func$ in program $\prog$ and the fault location $\lineid$, we generate a sketch by replacing the line $\lineid$ with a hole and complete the sketch based on the given tests.
As shown in Algorithm~\ref{algo:synthesis}, our patch synthesis algorithm consists of three steps: (1) introducing a hole at the fault location of program $\prog$ to obtain a sketch $\sketch$, (2) generating a context-free grammar $\grammar$ to capture the search space for the expression to fill in the hole, and (3) completing the sketch $\sketch$ by finding a correct expression accepted by $\grammar$.
In what follows, we describe each of the steps in more details.

\begin{figure}[!t]
\begin{algorithm}[H]
\caption{Patch Synthesis}
\label{algo:synthesis}
\begin{algorithmic}[1]
\Procedure{\textsc{SynthesizePatch}}{$\prog, \exs, \func, \lineid$}
\vspace{2pt}
\small
\Statex \textbf{Input:} Program $\prog$, examples $\exs$, faulty function $\func$, faulty line $\lineid$
\Statex \textbf{Output:} Repaired program $\prog'$ or $\bot$ to indicate failure

\State $\sketch \gets \textsf{IntroduceHole}(\prog, \func, \lineid)$;
\State $\grammar \gets \textsf{GenerateGrammar}(\sketch)$;
\State \Return $\textsc{CompleteSketch}(\sketch, \grammar, \exs)$;
\EndProcedure
\end{algorithmic}
\end{algorithm}
\vspace{-10pt}
\end{figure}

\vspace{5pt}
\noindent \textbf{Hole introduction.}
To generate a sketch from the original program and target function $\func$, we replace the \emph{maximal} expressions at the fault location in $\func$ with a hole. The maximal expressions to be considered are determined by the kind of the faulty statement. In particular, we introduce holes for the right-hand-side expressions of assignments, conditional expressions of jump statements, return values of return statements, and functions and arguments for function invocations. Replacing these expressions with holes turns the original program into a sketch and reduces the repair problem into a problem that aims to find a correct expression to instantiate the hole.

\vspace{5pt}
\noindent \textbf{Search space generation.}
After the hole is generated in the sketch, we still need to determine the search space for candidate expressions that can potentially result in a correct patch. The key challenge for defining the search space is to ensure the search space indeed includes the correct patch expression.
To address this challenge, we define a context-free grammar as shown in Figure~\ref{fig:sketch-grammar}, which includes all expressions that are constructed using constants, variables, field accesses, function invocations, unary and binary operators.
While it is possible to obtain a fixed grammar that contains a comprehensive set of constructs that are useful for many network programs, the search space of such grammar may become unnecessarily large for a particular program. To solve this problem, we parameterize all constants, variables, fields, functions, and operators over the sketch and only instantiate constructs that are in scope. For example, given a particular sketch with a hole, we only populate the variable set with all local and global variables that are in scope of the hole. As another example, if the hole corresponds to the conditional expression of a if statement, we only add logical operators to the grammar.

\begin{figure}[!t]
\centering
\[
\begin{array}{rcl}
\emph{Expr} &::=& \emph{UnaryOp}(\emph{Expr}) \ \ | \ \ \emph{BinaryOp}(\emph{Expr}, \emph{Expr}) \\
            & | & v.f(\emph{Expr}, \ldots, \emph{Expr}) \ \ | \ \ v.\emph{Expr} \ \ | \ \ v \ \ | \ \ c \\
\end{array}
\]
\[
\begin{array}{c}
v \in \textbf{Variables} \quad c \in \textbf{Constants} \quad f \in \textbf{Functions}
\end{array}
\]
\caption{Grammar for completions of sketch holes.}
\label{fig:sketch-grammar}
\end{figure}

\begin{example}
Consider again the \textrm{isSameAs} function from our motivating example in Figure~\ref{fig:example}.
\begin{lstlisting}[language=java]
public class FirewallRule {
  public MacAddress dl_dst; public boolean any_dl_dst;
  public boolean isSameAs(FirewallRule r) {
    if (... || any_dl_dst != r.any_dl_dst
            || (any_dl_dst == false &&
                dl_dst != r.dl_dst))
        return false;
    return true;
  }
}
\end{lstlisting}
Suppose the fault localization procedure finds the fault location is Line 6, we obtain the sketch
\begin{lstlisting}[language=java, numbers=none]
  public boolean isSameAs(FirewallRule r) {
    if (... || any_dl_dst != r.any_dl_dst
            || (any_dl_dst == false && ?? ))
        return false;
    return true;
  }
\end{lstlisting}
where $\hole$ denotes a hole in the generated sketch. The search space of expressions for filling in the hole $\hole$ can be described by the following context-free grammar with start symbol $\emph{Expr}$
\[
\begin{array}{rcl}
\emph{Expr} &::=& \emph{true} \ \ | \ \ \emph{false} \ \ | \ \ \emph{any\_dl\_dst} \ \ | \ \ r.\emph{any\_dl\_dst} \\
            & | & \emph{any\_dl\_dst} == r.\emph{any\_dl\_dst} \ \ | \ \ \emph{any\_dl\_dst}.\emph{equals}(r.\emph{any\_dl\_dst}) \\
            & | & !\emph{Expr} \ \ | \ \ \emph{Expr} \&\& \emph{Expr} \ \ | \ \ Expr || Expr \\
\end{array}
\]
\end{example}

\vspace{5pt}
\noindent \textbf{Sketch completion.}
Given a program sketch $\sketch$ with a context-free grammar $\grammar$ for its hole, the goal of sketch completion is to find an expression accepted by $\grammar$ such that the program obtained by replacing the hole with this expression can pass the given tests $\exs$. To solve the sketch completion problem, we use a top-down synthesis approach and perform depth-first search in the space of expressions generated by the grammar $\grammar$ to find the correct expression. The algorithm is summarized in Algorithm~\ref{algo:completion}.

\begin{figure}[!t]
\begin{algorithm}[H]
\caption{Sketch Completion}
\label{algo:completion}
\begin{algorithmic}[1]
\Procedure{\textsc{CompleteSketch}}{$\sketch, \grammar, \exs$}
\vspace{2pt}
\small
\Statex \textbf{Input:} Sketch $\sketch$, grammar $\grammar$, examples $\exs$
\Statex \textbf{Hyperparameter:} Maximum number of expansions $K$
\Statex \textbf{Output:} Completed sketch $\sketch'$ or $\bot$ to indicate failure

\If{$\textsf{ExpansionNum}(\sketch, \grammar) > K$} \Return $\bot$; \EndIf
\If{$\textsf{IsCompleteProgram}(\sketch)$}
    \If{$\textsf{Verify}(\sketch, \exs)$} \Return $\sketch$;
    \Else \ \Return $\bot$; \EndIf
\EndIf
\State $N \gets \textsf{GetANonTerminal}(\sketch)$;
\For{\textbf{each} production $\alpha ::= \beta \in \textsf{Productions}(\grammar)$}
    \If{$\alpha = N$}
        \State $\sketch' \gets \textsf{Expand}(\sketch, N, \beta)$;
        \State $\sketch^* \gets \textsc{CompleteSketch}(\sketch', \grammar, \exs)$;
        \If{$\sketch^* \neq \bot$} \Return $\sketch^*$; \EndIf
    \EndIf
\EndFor
\State \Return $\bot$;
\EndProcedure
\end{algorithmic}
\end{algorithm}
\vspace{-10pt}
\end{figure}

At a high level, the \textsc{CompleteSketch} procedure in Algorithm~\ref{algo:completion} starts with a sketch obtained by replacing the hole with the root non-terminal of grammar $\grammar$. The procedure progressively expands the non-terminal based on productions in $\grammar$ until it finds a sketch without any non-terminal (i.e., a complete program) that can pass the tests $\exs$.
Since there could be recursive production rules that make infinite number of candidate programs accepted by the grammar, the sketch completion procedure may not terminate. To resolve this issue, we count the number of expansions used to obtain a sketch or complete program and put a limit on the maximum number of expansions. In this way, the \textsc{CompleteSketch} procedure is guaranteed to terminate in finite time.

As shown in Algorithm~\ref{algo:completion}, the \textsc{CompleteSketch} procedure first checks the current number of expansions and immediately returns $\bot$ if the number exceeds the predefined hyper-parameter $K$ (Line 2). Next, it checks the termination condition of the recursion (Line 3), i.e. whether $\sketch$ is a complete program. If  $\sketch$ is complete or does not contain any non-terminal, \textsc{CompleteSketch} executes the tests $\exs$ on $\sketch$. If $\sketch$ indeed passes the tests $\exs$, sketch completion succeeds with $\sketch$ (Line 4); otherwise, the procedure returns $\bot$ indicating failure (Line 5).
If sketch $\sketch$ has at least one non-terminal symbol, \textsc{CompleteSketch} obtains the next non-terminal symbol $N$ to expand (Line 6). Then it enters a loop (Lines 7 -- 11) and enumerates all productions in $\grammar$ where the left-hand side of the production is $N$. In particular, for each production $N ::= \beta$, \textsc{CompleteSketch} obtains a new sketch $\sketch'$ from $\sketch$ by replacing the non-terminal $N$ with $\beta$ (Line 9) and recursively invokes itself with the new sketch $\sketch'$ (Line 10). If a correct completion $\sketch^*$ is found by the recursive call, then the caller \textsc{CompleteSketch} also returns $\sketch^*$ (Line 11); otherwise, it moves on to the next production. This process is repeated until all productions in $\grammar$ are checked. If no correct completion exists, \textsc{CompleteSketch} returns $\bot$ to indicate failure (Line 12).

\begin{theorem}[Soundness of sketch completion] \label{thm:synth-soundness}
Given a sketch $\sketch$, a grammar $\grammar$ for expressions to fill in the hole in $\sketch$, and a set of unit tests $\exs$, let $\sketch'$ be the return value of $\textsc{CompleteSketch}(\sketch, \grammar, \exs)$. If $\sketch' \neq \bot$, then $\sketch'$ can pass all unit tests in $\exs$.
\end{theorem}
\begin{proof}
See Appendix A.
\end{proof}

\begin{theorem}[Completeness of sketch completion] \label{thm:synth-completeness}
Given a sketch $\sketch$, a grammar $\grammar$, a set of unit tests $\exs$, and a hyper-parameter $K$, if $\textsc{CompleteSketch}(\sketch, \grammar, \exs) = \bot$, then there does not exist an expression $e$ accepted by $\grammar$ such that (1) the number of expansions from the root symbol of $\grammar$ to $e$ is no more than $K$, and (2) substituting the hole in $\sketch$ with $e$ results in a program that passes all unit tests in $\exs$.
\end{theorem}
\begin{proof}
See Appendix A.
\end{proof}

\section{Implementation} \label{sec:impl}

We have implemented the proposed repair technique in a tool called \toolname. \toolname leverages the Soot static analysis framework~\cite{soot-cetus11} to convert Java programs into Jimple code, which provides a succinct yet expressive set of instructions for analysis. In addition, \toolname utilizes the Rosette tool~\cite{rosette-pldi14} to perform symbolic reasoning for fault localization and patch synthesis.
While our implementation closely follows the algorithm presented in Section~\ref{sec:related}, we also conduct several optimizations that are important to improve the performance of \toolname.

\vspace{5pt}
\noindent \textbf{Validating patches with local specifications.}
Observe that the patch synthesis procedure could potentially validate a large number of candidate patches, it is crucial to optimize the validation procedure of \toolname for better performance. Our key idea is to validate the correctness of a candidate patch based on the pre- and post-condition of fault locations, rather than executing the tests from the beginning of the program. Specifically, for each provided unit test, \toolname symbolically executes the network program and infers the pre- and post-states of the faulty line in the process of fault localization. Then in the patch synthesis phase, \toolname can execute each candidate patch from the inferred pre-state and check if the execution result is consistent with the inferred post-state. This validation procedure enables fast checking of each candidate patch, because it avoids repeated symbolic execution of correct statements and only executes those patched statements.

\vspace{5pt}
\noindent \textbf{Memories for different types.}
Since the conversion between bitvectors and integers imposes significant overhead on running time, \toolname divides the memory into two parts, one for integers and the other for bitvectors. In this design, \toolname automatically selects the memory chunk based on the variable types. In particular, it only stores integer values in the integer memory and likewise bitvectors in the bitvector memory. This optimization significantly improves the performance of symbolic reasoning, because there is no need to convert between different data types.

\vspace{5pt}
\noindent \textbf{Stack and heap.}
In order to reduce the number of memory operations, \toolname also divides the memory into stack and heap. As is standard, stack only stores static data and its layout is deterministic. For example, the locations of function arguments and return values are fixed and statically available, so they are stored on stack. Therefore, stacks are implemented using fixed-size vectors, and thus can be efficiently accessed for read and write operations.
On the other hand, heap stores dynamic data that are usually not known at compile time, such as allocated objects. Since the heap size cannot be determined beforehand, \toolname uses an uninterpreted function $f(x)$ to represent heaps, where $x$ is the address and $f(x)$ is the value stored at $x$.

\vspace{5pt}
\noindent \textbf{String values.}
Since reasoning over string values is a challenging task and not always necessary for repairing network programs, we simplified the representation of strings with integer values. Specifically, \toolname maps each string literal to a unique integer and represent all string operations (e.g. concatenation) with uninterpreted functions. While many existing techniques~\cite{string-cav14,string-fmsd17} can improve the precision of string analysis, we find our current approximation is sufficient for repairing network programs in our experimental evaluation.


\vspace{5pt}
\noindent \textbf{Bounded program analysis.}
In order to improve the repair time, \toolname only performs bounded program analysis for fault localization and patch synthesis. Namely, we unroll loops and inline functions up to $K$ times, where $K$ is a predefined hyper-parameter. In this way, function summaries can be easily and efficiently computed using symbolic execution. While it is possible to incorporate invariant inference and recursive function summarization techniques, we do not implement them in the current version of \toolname and leave it as future work.


\section{Evaluation} \label{sec:eval}

To evaluate the proposed techniques, we perform experiments that are designed to answer the following research questions:
\begin{itemize}
\item[\textbf{RQ1}] Is \toolname effective to repair realistic network programs?
\item[\textbf{RQ2}] How efficient are the fault localization and repair techniques in \toolname?
\item[\textbf{RQ3}] How helpful are modular analysis and domain-specific abstraction for repairing network programs?
\item[\textbf{RQ4}] How does \toolname perform compared to other repair tools for Java programs?
\end{itemize}

\paragraph{Benchmark collection}
To obtain realistic benchmarks, we crawl the commit history of Floodlight~\cite{floodlight-web21}, an open-source SDN controller that supports the OpenFlow protocol, and identify commits based on the following criteria:
\begin{enumerate}
\item The commit message contains keywords about repairing bugs, e.g., ``bug'', ``error'', ``fix'';
\item The commit changes no more than three lines of code.
\end{enumerate}
These criteria are important because they are able to distinguish commits caused by bug repairs from those generated for non-repair scenarios, such as code refactoring, adding functionalities, etc.
Following these criteria, we have collected 10 commits from the Floodlight repository and adapted them into our benchmarks. Specifically, given a commit in the repository, we take the code before the commit as the faulty network program and the version after the commit as the ground-truth repaired program. The code is post-processed and the parts irrelevant to the bug of interest are removed. We also identify corresponding unit tests and modify them to directly reveal the bug as appropriate. Each benchmark in our evaluation consists of a faulty network program and its corresponding unit tests.

\paragraph{Experimental setup}
All experiments are conducted on a computer with 4-core 2.80GHz CPU and 16GB of physical memory, running the Arch Linux Operating system. We use Racket v7.7 as the compiler and runtime system of \toolname and set a time limit of 1 hour for each benchmark.

\subsection{Main Results}

\begin{table}[!t]
\small
\centering
\vspace{10pt}
\begin{tabular}{clcccccccc}
\hline\hline
\multirow{2}{*}{\textbf{ID}} &
\multirow{2}{*}{\textbf{Module}} &
\multirow{2}{*}{\textbf{LOC}} &
\multirow{2}{*}{\textbf{\# Funcs}} &
\multirow{2}{*}{\textbf{\# Tests}} &
\multirow{2}{*}{\textbf{Succ}} &
\multirow{2}{*}{\textbf{Exp}} &
\textbf{Loc} &
\textbf{Synth} &
\textbf{Total} \\
& & & & & & & \textbf{Time (s)} & \textbf{Time (s)} & \textbf{Time (s)} \\
\hline
1 & DHCP & 212 & 17 & 2 & Yes & Yes & 40 & 117 & 157 \\
2 & Load Balancer & 336 & 28 & 2 & No & No & - & - & - \\
3 & Firewall & 262 & 13 & 2 & Yes & Yes & 893 & 197 & 1090 \\
4 & DHCP & 431 & 32 & 2 & Yes & Yes & 95 & 39 & 134 \\
5 & Utility & 809 & 65 & 2 & No & No & - & - & - \\
6 & Routing & 605 & 44 & 3 & Yes & Yes & 271 & 179 & 450 \\
7 & Utility & 454 & 45 & 2 & Yes & Yes & 39 & 46 & 85 \\
8 & Learning Switch & 738 & 34 & 2 & Yes & No & 571 & 595 & 1166 \\
9 & Database & 442 & 17 & 2 & Yes & No & 310 & 2139 & 2449 \\
10 & Link Discovery & 671 & 46 & 2 & Yes & No & 268 & 158 & 426 \\
\hline\hline
\end{tabular}
\vspace{5pt}
\caption{Experimental results of \toolname.}
\label{tab:results}
\end{table}

Our main experimental results are summarized in Table~\ref{tab:results}. The column labeled ``Module'' in the table describes the network module to which the benchmark belong. The next two columns labeled ``LOC'' and ``\# Funcs'' show the number of lines of source code (in Jimple) and the number of functions, respectively. The ``\# Tests'' column presents the number of unit tests used for fault localization and patch synthesis. Next, the ``Succ'' and ``Exp'' columns show whether \toolname can successfully repair the program and if the generated patch is exactly the same as the ground-truth. Since \toolname returns the first fix that can pass all provided test cases, the repaired programs are not necessarily the same as those expected in the ground-truth. In this case, the table will show a ``Yes'' in the ``Succ'' column and a ``No'' in the ``Exp'' column. Finally, the last three columns in Table~\ref{tab:results} denote the fault localization time, patch synthesis time and the total running time of \toolname.

As shown in Table~\ref{tab:results}, there is a range of 13 to 65 functions in each benchmark and the average number of functions is 34 across all benchmarks. Each benchmark has 212 -- 809 lines of Jimple code, with the average being 496. \toolname succeeds in repairing 8 out of 10 benchmarks. Furthermore, for 5 benchmarks that can be successfully repaired, \toolname is able to generate exactly the same fix as ground-truth. Given that our benchmarks cover programs from a variety of modules of Floodlight, such as DHCP Server, Firewall, etc, we believe that \toolname is effective to repair realistic network programs (RQ1).

To understand why \toolname is not able to repair benchmark 2 and 5, we manually inspect the corresponding network programs and the execution logs. We found \toolname failed to localize the fault of benchmark 2 due to its incomplete support for the hash map data structure. Ideally, the hash map should be modeled as an unbounded data structure with dynamic allocation, which is beyond the capability of our current symbolic analysis. For Benchmark 5, \toolname is able to localize the fault but not able to synthesize the correct patch. The expected patch for benchmark 5 requires replacing an invocation of a function with side effects with another function, which is out of the ability of \toolname's patch synthesizer.

Regarding the efficiency, \toolname can repair 8 benchmarks in an average of 744 seconds with only 2 to 3 test cases. The fault localization time ranges from 39 seconds to 893 seconds, with 50\% of the benchmarks within five minutes. The patch synthesis time ranges from 39 seconds to 2139 seconds, with 60\% of the benchmarks within five minutes. In summary, the evaluation results show that \toolname only takes minutes to localize bugs in a faulty program and synthesize a correct patch based on two to three unit tests (RQ2).

\subsection{Ablation Study}

To explore the impact of modular analysis and domain-specific abstraction on the proposed repair technique, we develop three variants of \toolname:
\begin{itemize}[leftmargin=*]
\item \toolname-\textsc{NoMod} is a variant of \toolname without modular analysis. Specifically, given a faulty network program $P$, \toolname-\textsc{NoMod} inlines the functions in $P$ but still uses abstractions for network data structures for fault localization and patch synthesis. It does not compute or reuse function summaries for symbolic reasoning.
\item \toolname-\textsc{NoAbs} is a variant of \toolname without domain-specific abstraction. In particular, \toolname-\textsc{NoAbs} does not use abstractions for any function in network data structures. Instead, it uses the original concrete implementation of those functions for symbolic reasoning. If the implementation is written in a different language, we manually translate the implementation to Java. 
\item \toolname-\textsc{NoModAbs} is a variant of \toolname without modular analysis or domain-specific abstraction. Essentially, \toolname-\textsc{NoModAbs} simply inlines all functions in the faulty program, including those functions in the network data structures, and performs symbolic analysis for fault localization and patch synthesis.
\end{itemize}

\begin{figure}[!t]
\centering
\begin{tikzpicture}[scale = 0.85]
\begin{axis}[
    width=5.1in,
    height=3.0in,
    legend cell align=left,
    legend entries={\toolname, \toolname-\textsc{Abs}, \toolname-\textsc{Mod}, \toolname-\textsc{ModAbs}},
    legend style={at={(0,1.0)}, font=\scriptsize, legend columns=1, anchor=north west},
    x label style={below=0pt},
    y label style={below=-5pt},
    xmin=0,
    ymin=0,
	xlabel={\# Solved Benchmarks},
	ylabel={Running Time (s)},
	cycle list name=exotic,
	every axis plot/.append style={ultra thick}
]
\addplot+[green,mark=*] table {orion.data};
\addplot+[brown,mark=square*] table {no-abs.data};
\addplot+[blue,mark=diamond*] table {no-mod.data};
\addplot+[red,mark=triangle*] table {no-modabs.data};
\end{axis}
\end{tikzpicture}
\caption{Comparing \toolname against three variants.}
\label{fig:ablation}
\end{figure}
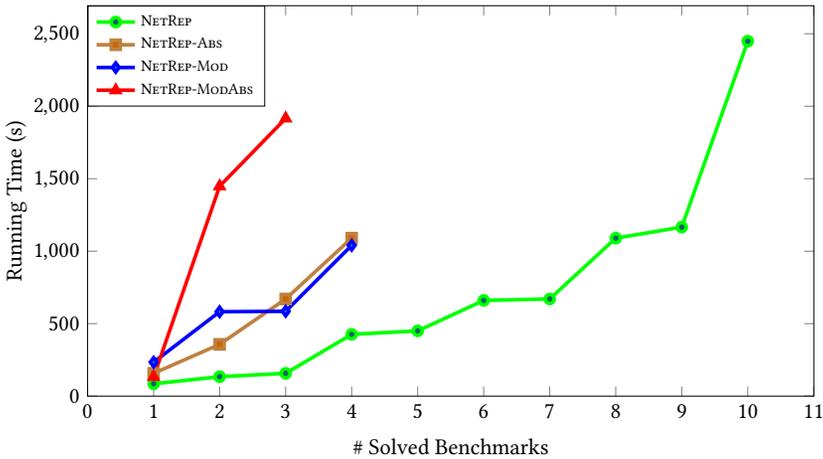

To understand the impact of modular analysis and domain-specific abstraction, we run all variants on the 10 collected benchmarks. For each variant, we measure the total running time (including time for fault localization and time for patch synthesis) on each benchmark, and order the results by running time in increasing order. The results for all variants are depicted in Figure~\ref{fig:ablation}. All lines stop at the last benchmark that the corresponding variant can solve within 1 hour time limit.

As shown in Figure~\ref{fig:ablation}, \toolname-\textsc{NoAbs} can only solve 4 out of 10 
benchmarks in the evaluation, with the average running time being 569 seconds. Similarly,
\toolname-\textsc{NoMod} is only able to solve 4 out of 10 benchmarks within the 1 hour 
time limit, and the average running time is 610 seconds. Among all different variants,
\toolname-\textsc{NoModAbs} solves the least number of benchmarks: 3 out of 10. 
For the ones that it can solve, the average running time is 1165 seconds.
This experiment shows that modular analysis and domain-specific abstraction are both great
boost to \toolname's efficiency to repair network programs (RQ3).

\subsection{Comparison with the Baseline}

To understand how \toolname performs compared to other Java program repair tools, we compare \toolname against a state-of-the-art tool called \jaid~\cite{jaid-ase17} on our benchmarks. Specifically, \jaid takes as input a faulty Java program, a set of unit tests, and a function signature for fault localization and patch synthesis. Note that \jaid solves a simpler repair problem than \toolname, because it requires the user to specify a function that is potentially incorrect in the program, whereas \toolname does not need input other than the faulty program and unit tests. In order to run \jaid on our benchmarks, we adjust the format of our benchmarks into \jaid's format and provide the faulty function (known from the ground truth) as input for \jaid.

\begin{table}[!t]
\small
\centering
\begin{tabular}{ccccc}
\hline\hline
\textbf{Tool} &
\textbf{Expected} &
\textbf{Succeed but unexpected} &
\textbf{Failed} &
\textbf{Total} \\
\hline
\toolname & 5 & 3 & 2 & 10 \\
\jaid & 2 & 6 & 2 & 10 \\
\hline\hline
\end{tabular}
\vspace{5pt}
\caption{Comparison between \toolname and \jaid.}
\label{tab:jaid}
\end{table}

The comparison results are summarized in Table~\ref{tab:jaid}. As we can see from the table,
\jaid is able to generate valid fixes with respect to the test cases for 8 out of 10 benchmarks.
But after a manual inspection, we find that only 2 of them are the same as those shown in the ground-truth. For the remaining two benchmarks, \jaid fails to generate a valid patch. In particular, \jaid exceeds the time limit for one benchmark and runs out of memory for the other.

It is not feasible to reasonably compare the running time between \jaid and \toolname, 
because \jaid is not designed to stop after finding the first valid fix. Instead, it will
generate a large number of candidate patches and output a ranked list of valid ones
among them, which takes excessively long to eventually finish. 

\toolname outperforms \jaid in terms of repairing accuracy. In particular, \toolname is able to repair the same number of benchmarks as \jaid and find the expected fix among five of them, whereas \jaid is only
able to find the expected fix on two.

In addition, \toolname outputs the first valid patch it finds, while \jaid may produce hundreds or thousands of candidate patches, which requires extra ranking heuristics. This difference can be explained by the amount of semantic information used by each tool.
Specifically, \jaid monitors a selected set of states chosen by dynamic semantic analysis, and localizes
potential bug locations by a heuristic-based ranking algorithm over the values of states collected through the execution of test cases. Like similar preceding systems, this method 
relies on matching the ranking algorithm's heuristics with specific tasks, 
as well as a number of test cases to generate enough state information to use. 
\toolname strictly encodes the semantic information of the entire program and infers the bug
location as well as the specification for the patch from this encoding. Therefore, \toolname is less
likely to overfit to specific test cases or algorithm heuristics.

In summary, \toolname is more effective in automatically fixing bugs in network programs
compared to state-of-the-art repairing tools for Java programs, especially with respect to repairing
accuracy and avoiding overfitting (RQ4).


\section{Related Work} \label{sec:related}

In this section, we survey related work in automated program repair, fault localization, patch synthesis, as well as verification, synthesis, and repair techniques for software-defined networking.

\paragraph{Automated program repair.}

Automated program repair is an active research area that aims to automatically fix the mistakes in programs based on specifications of correctness criteria~\cite{LeGoues-cacm19,dfix-pldi19,semcluster-pldi19,saver-icse20}, with a variety of applications such as aiding software development~\cite{sapfix-icse19}, finding security vulnerabilities~\cite{angelix-icse16}, and teaching novice programmers~\cite{Wang-pldi18,Gulwani-pldi18}. Different techniques have been proposed to solve the automated program repair problem, including heuristics-based techniques~\cite{Harman-cacm10,spr-fse15}, semantics-based techniques~\cite{angelix-icse16,s3-fse17}, and learning-based techniques~\cite{rite-pldi20,dlfix-icse20,prophet-popl16,codephage-pldi15}.
For example, \textsc{SPR}~\cite{spr-fse15} decomposes the repair problem into stages, e.g. transformation schema select, condition synthesis, and applies a set of heuristics to prioritize the order in which it validates the repair. \textsc{Prophet}~\cite{prophet-popl16} improves \textsc{SPR} by learning a probabilistic model from correct code to prioritize the search procedure. \textsc{Rite}~\cite{rite-pldi20} repairs type-error of OCaml programs by learning templates from training, predicting the template using classifiers, and synthesizing repairs based on enumeration and ranking. \textsc{DLFix}~\cite{dlfix-icse20} reduces the automated repair problem into a code transformation learning problem and proposes a two-layer deep model to learn transformations from prior bug fixes.
\toolname is related to this line of research, and it is a semantics-based automated repair tool. Different from prior work, \toolname is specialized to repair network programs based on modular analysis and network data structure abstractions.

\paragraph{Fault localization.}
One of the most important techniques related to automated program repair is fault localization, which studies the problem of finding which part of program is incorrect according to the specification. Researchers have developed various approaches to fault localization, including spectrum-based, learning-based, and constraint-based techniques.
Specifically, the spectrum-based techniques~\cite{s3-fse17,spectrum-ase09,spectrum-taicpart07,ample-ecoop05,Renieris-ase03,pinpoint-dsn02,tarantula-icse02} perform fault localization by identifying which part of program is active during a run through execution profiles (called program spectrum). For example, \textsc{Ample}~\cite{ample-ecoop05} identifies fault locations for object-oriented programs by analyzing sequences of method calls.
Learning-based techniques~\cite{deepfl-issta19,multric-icsme14,learning-ieice17} typically train machine learning models to predict and rank possible fault locations. For instance, \textsc{DeepFL} proposes a deep learning approach that automatically learns existing or latent features for fault localization via learning-by-rank.
By contrast, constraint-based techniques~\cite{Jose-pldi11,bugassist-cav11} encode the semantics of problems as logical constraints and reduce the fault localization problem into constraint satisfaction problem. Jose and Majumdar~\cite{Jose-pldi11} propose to encode the execution trace of the failing test as a trace formula and identify potential errors by finding a maximal set of clauses in the formula.
In spirit, \toolname uses a similar idea for fault localization. However, \toolname has two important differences from the techniques proposed in prior work by Jose and Majumdar. First of all, \toolname does not solve maximum satisfiability problem but rather forces to find a solution that can satisfy $N-K$ clauses for a $N$-clause encoding. It is helpful to avoid solving the difficult MaxSMT problem. Second, \toolname performs modular analysis and enables localizing faults involving function calls, whereas the prior work mainly focuses formulas obtained from execution traces.

\paragraph{Patch synthesis.}
Another important technique related to automated program repair is patch synthesis, which concerns how to generate a correct patch given the correctness specification and a fault location. Many synthesis algorithms have been developed for generating patches, including enumerative search~\cite{s3-fse17}, constraint-based techniques~\cite{angelix-icse16}, statistical model~\cite{acs-icse17}, machine learning~\cite{deepfix-aaai17}, and so on.
For example, \textsc{Angelix}~\cite{angelix-icse16} performs symbolic execution to obtain angelic forests, which contains enough semantic information to guide multi-line patch synthesis. ACS~\cite{acs-icse17} combines the heuristic ranking techniques based on the code structure, documentation of programs, and expressions in existing projects to automatically generate precise conditional expressions. DeepFix~\cite{deepfix-aaai17} trains a multi-layered sequence-to-sequence neural network to predict fault locations and corresponding patches. 
In terms of patch synthesis, \toolname generates a context-free grammar from the context of fault locations and performs enumerative search based on the grammar to synthesize patches. It does not require machine learning model or statistical information for ranking all possible patches. However, it is conceivable that \toolname will benefit from the guidance of such ranking techniques.

\paragraph{Verification and synthesis for SDN}
In the networking domain, several verification tools~\cite{vericon-pldi14,nod-nsdi15,veriflow-nsdi13,kinetic-nsdi15} have been proposed based on either model checking or theorem proving. For example, \textsc{VeriCon}~\cite{vericon-pldi14} utilizes first-order logic to describe network topologies and invariants. It performs deductive verification to verify the correctness of SDN programs on all admissible topologies and for all possible sequences of network events. \textsc{Kinetic}~\cite{kinetic-nsdi15} provides a domain-specific language that enables network operators to control networks in a dynamic way. It also automatically verifies the correctness of control programs based on user-provided temporal properties.
In addition to verification, synthesis techniques~\cite{McClurg-pldi15,McClurg-cav17,Padon-pldi15} have also been proposed to aid software-defined networking. For instance, McClurg et al.~\cite{McClurg-cav17,McClurg-pldi15} proposes synthesis techniques for updating global configurations and generating synchronizations for distributed controllers. Padon et al.~\cite{Padon-pldi15} formalizes the correctness and optimality requirements of decentralizing network policies and identifies a class that is amenable to synthesis of optimal rule installation policies.
By contrast, \toolname aims to repair network programs automatically, which is a different problem than SDN verification or synthesis.

\paragraph{Repair for network programs.}
Our work is most related to automated repair of network programs in the SDN domain~\cite{MetaProvenance-hotnets15,MetaProvenance-nsdi17,McClurg-fmcad16}. Prior work about auto-repair~\cite{MetaProvenance-hotnets15,MetaProvenance-nsdi17} relies on using Datalog to capture the operational semantics of the target language to be repaired. The repair techniques work for domain specific languages (e.g. Datalog or Ruby on Rails) with simple structure. Similarly, Hojjat et al.~\cite{McClurg-fmcad16} proposes a framework based on horn clause repair problem to help network operators fix buggy configurations. However, \toolname targets Java network programs with object-oriented features and more complex constructs, which cannot be handled by existing techniques.


\section{Limitations and Future Work} \label{sec:limitations}

In this section, we discuss several limitations of \toolname that we plan to improve in future work. 

First, \toolname repairs the faulty network program with the first correct patch that can pass all provided unit tests. However, the repaired program may be different from the fix intended by developers. To address this problem and find the user-intended fix, we plan to introduce user interaction in future work. In particular, the synthesis procedure will not stop after finding the first correct patch. Instead, it enters an interactive loop that asks for feedback from developers and search for the next correct patch until developers are satisfied with the repaired program.

Second, \toolname focuses on local changes of the program for repair. Specifically, \toolname only makes local changes around the lines determined as fault locations, such as replacing an expression or statement with another, using a different function or operator, etc. Thus, \toolname may not be able to repair network programs that require complicated changes of control flow structures. We envision this limitation to be overcome by a more sophisticated patch synthesis techniques such as defining a domain-specific language for edits (e.g. operators for introducing branches of conditional statements) and search over the DSL to find the corresponding repair.

Third, in order to force symbolic execution to terminate in finite time, \toolname currently unrolls all loops in the network program, which may cause the fault localization technique to miss a potential fault. To systematically eliminate this limitation, we plan to leverage invariant inference techniques to generate loop invariants for programs to repair, so that the symbolic executor can directly use invariants for loops while ensuring the termination.

\section{Conclusion} \label{sec:concl}

In this paper, we have proposed an automated repair technique for network programs. The repair technique takes as input a faulty network program and a suite of unit tests that can reveal the fault. It produces as output a repaired network program such that the program can pass all unit tests. Our technique internally performs symbolic reasoning for fault localization and patch synthesis. To improve symbolic reasoning, it combines domain-specific abstractions for network data structures and modular analysis to facilitate reuse of function summaries. To evaluate the proposed repair technique, we have implemented a tool called \toolname and evaluated it on 10 benchmarks adapted from the Floodlight framework. The experimental results demonstrate that \toolname is effective for repairing realistic network programs.

\bibliography{references}

\ifextended{
\appendix
\section{Proofs} \label{sec:proofs}

\begin{proof}[Proof of Theorem~\ref{thm:synth-soundness}]
Observe that the \textsc{CompleteSketch} in Algorithm~\ref{algo:completion} only returns a non-$\bot$ value if \textsf{Verify}($\sketch, \exs$) returns true (Line 4). If the return value $\sketch'$ of \textsc{CompleteSketch} is not $\bot$, it holds that $\sketch'$ can pass all unit tests in $\exs$ by the soundness of \textsf{Verify}.
\end{proof}

\begin{proof}[Proof of Theorem~\ref{thm:synth-completeness}]
Proof by induction on the maximum number of expansions $K$.
\begin{itemize}
\item Base case: $K = 0$.
The theorem holds trivially because no expression $e$ is accepted by $\grammar$ if no expansion is allowed.
\item Inductive case:
Suppose there does not exist an expression $e$ that satisfies all conditions for $K = n$, we need to prove there is no expression $e$ satisfying all conditions for $K = n + 1$.
Let us discuss by two cases.
\begin{enumerate}
    \item $\sketch$ does not have any non-terminal after $n$ expansions, i.e. $\sketch$ is a complete program. By inductive hypothesis, there does not exist an expression $e$ such that $e$ requires no more than $K$ expansions from the root of $\grammar$ and $e$ can lead to an instantiation that passes all tests in $\exs$. Since $\sketch$ does not have non-terminals, increasing the maximum number of expansions does not change the execution of the \textsc{CompleteSketch} at all. Thus, there does not an expression $e$ that satisfies all the conditions for $K = n + 1$.
    \item $\sketch$ still has non-terminals after $n$ expansions. Without loss of generality, let us assume the next non-terminal in $\sketch$ is $N$. Given that \textsc{CompleteSketch}($\sketch, \grammar, \exs$) = $\bot$, so the procedure must return at Line 2, Line 5, or Line 12. In addition, consider that we have increased the maximum number of expansion by 1 and $sketch$ is not a complete program, we know the return location is not Line 2 or 5. Thus the only possible return location is Line 12 and it holds that all $\sketch^*$ at Line 10 are $\bot$. By inductive hypothesis, all sketches from the previous recursive call cannot pass all the test cases in $\exs$, so there does not exists an expression $e$ that requires no more than $n+1$ expansions will result in a correct instantiation of $\sketch$.
\end{enumerate}
\end{itemize}
\end{proof}

}\fi

\end{document}